\documentclass[]{emulateapj}
\usepackage{graphicx,amsmath,rotating}% Include figure files

\begin{document}

\title{Pencil-Beam Surveys for Trans-Neptunian Objects:\\ Limits on Distant Populations}
\author{\bf Alex H. Parker}
\affil{\emph{Department of Astronomy, University of Victoria}}
\email{alexhp@uvic.ca}
\and
\author{\bf JJ Kavelaars}
\affil{\emph{Herzberg Institute of Astrophysics, National Research Council of Canada}}

\slugcomment{Submitted for publication in ICARUS}

\begin{abstract}
Two populations of minor bodies in the outer Solar System remain particularly elusive: Scattered Disk objects and Sedna-like objects. These populations are important dynamical tracers, and understanding the details of their spatial- and size-distributions will enhance our understanding of the formation and on-going evolution of the Solar System. By using newly-derived limits on the maximum heliocentric distances that recent pencil-beam surveys for Trans-Neptunian Objects were sensitive to, we determine new upper limits on the total numbers of distant SDOs and Sedna-like objects. While generally consistent with populations estimated from wide-area surveys, we show that for magnitude-distribution slopes of $\alpha \gtrsim 0.7-1.0$, these pencil-beam surveys provide stronger upper limits than current estimates in literature.
\end{abstract}

\shorttitle{Limits on Distant TNOs}
\shortauthors{Parker \& Kavelaars}

\maketitle

\section{Introduction}

A number of deep, narrow-angle ``pencil-beam'' surveys for distant Solar System objects have been undertaken over the past two decades, first appearing in the literature in Tyson et al. (1992). These surveys avoid the issue of trailing losses in long exposures by taking a large number of shorter exposures, predicting the sky rate of motion of sources of interest, then compensating for this motion in software before finally stacking the images. Due to the large number of rates at which images must be combined in order to properly compensate for the range of motions real objects can have, the orbital range over which this method is applied is limited to maintain computational feasibility. 

Parker \& Kavelaars (2010 submitted; hereafter P10) re-characterize the orbital limits of several published pencil-beam surveys and show that these orbital limits are poorly characterized in literature. As the re-derived maximum heliocentric distances these surveys were sensitive to ranges from $150-400$ AU, we find that these surveys were sensitive to several dynamically interesting yet currently poorly constrained populations; namely, Scattered Disk Objects (SDOs, eg., Trujillo et al. 2000; hereafter T00) and Sedna-like objects (SLOs, eg., Schwamb et al., 2009; hereafter S09). SDOs are highly-eccentric, non-resonant objects that have perihelia that interact with Neptune, and may be the source of Jupiter-family comets (eg., Duncan \& Levison 1997). SLOs are long-period objects that have perihelia high enough ($> 70$ AU) that they are out of reach of the giant planets but aphelia not distant enough for galactic tides to have significant effect on their orbits. Their current orbits require emplacement mechanisms that may be linked to dynamics of the Solar birth cluster (eg., Brasser et al. 2006), interactions with ``rogue'' planets (eg., Gladman \& Chan 2006), or close stellar passages (eg., Kenyon \& Bromley 2004). Few objects (one, in the case of Sedna) of either population have been discovered, and our understanding of these populations is severely hampered by these limited samples. Here we use the newly-derived heliocentric distance limits from P10 in conjunction with a simple survey simulator in order to determine new limits on the SDO and SLO populations. 

\section{Characterization of Outer Limits of Previous Survey Grids}

%%% Table 1 here

%%% Table 1
\begin{table*}
\centering
\footnotesize
\begin{tabular}{lcccc}
\multicolumn{5}{c}{\bf Table 1}\\
\multicolumn{5}{c}{\bf Pencil-Beam Survey Characterization}\\
\hline
\hline
 & $ d_{max}\,(i=0^\circ)$ & $d_{max}\,(i=180^\circ) $&$R_{50}$& $\Omega$\\
Survey & (AU) & (AU) & (mag) & (deg$^2$) \\
\hline
Fraser et al. 2008 & 164 & 184 & 25.25$^a$ & 3 \\
Fraser \& Kavelaars 2009 & 360 & 390 & 26.76 &0.255 \\
Fuentes et al. 2009 & 220 & 245 & 26.86 &0.255 \\
\hline
\hline
Bernstein et al. 2004 & \textit{1000}$^b$ & NA & 28.5 &0.019\\
\hline
 \end{tabular}\\
 {\footnotesize $^a$: Area-weighted mean $R_{50}$.\\
 $^b$: Not characterized by Parker \& Kavelaars 2010.}
 \end{table*}

In order to characterize the limits of parameter space searched by a survey, P10 generated a large sample of synthetic orbits that fall on the imaged field during the time of observations. These synthetic orbits spanned large ranges of heliocentric distance ($20-500$ AU), inclination ($0^\circ-180^\circ$), and eccentricity ($0-0.999$).  Ephemeris software was used to estimate the on-sky motion of each synthetic orbit generated, and then test the on-sky motion to ensure that it is within the ranges searched in the original survey. This characterization was performed for those surveys that included enough information in the literature to accurately reproduce the original search-grids.

Three surveys were characterized: Fraser et al. (2008), Fraser \& Kavelaars (2009), and Fuentes et al. (2009). Each survey was found to be sensitive to distances considerably greater than claimed, and this limit varied somewhat with inclination. Additionally, each survey was sensitive to inclinations as high as $180^\circ$, even though the target maximum inclinations were $70^\circ - 90^\circ$. Table 1 contains the derived limits and properties of each survey.

While P10 did not characterize Bernstein et al. (2004; hereafter B04), we include it for analysis here. The authors claim that their survey is sensitive to motions of sources at distances greater than 1000 AU and to inclinations as high as $45^\circ$. Because B04 imaged fields near quadrature (while the three surveys above were taken near opposition), it is not straightforward to apply the results of P10 to this survey. However, we are confident that adopting 1000 AU as a conservative distance limit for B04 is an adequate first-order estimate.

\section{Upper limits on distant populations from previous surveys}

\subsection{Survey Simulations}

%%% Figure 1
\begin{figure*}[t]
\begin{centering}
\includegraphics[width=10cm]{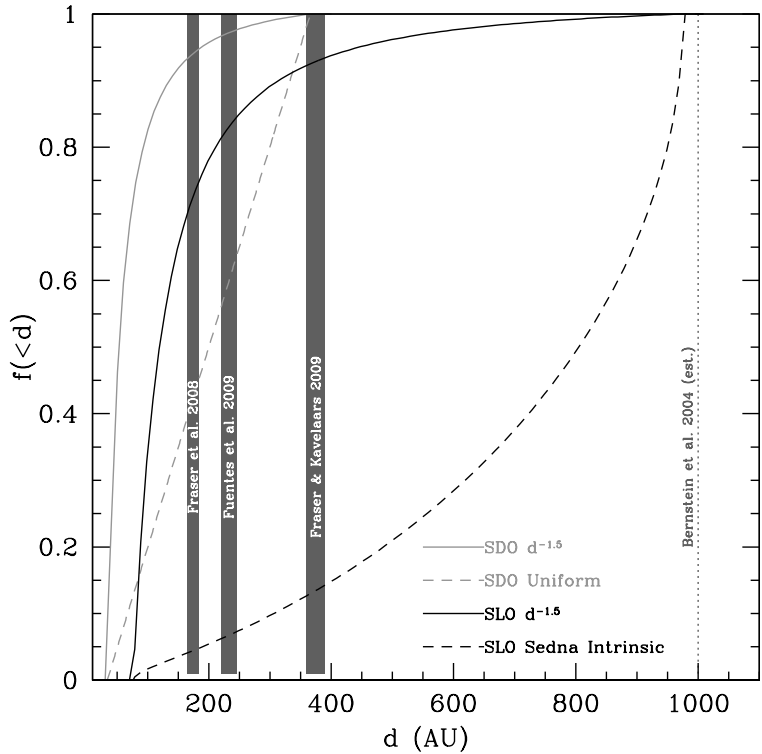}
\caption{Cumulative radial distributions and limits used in our simulations. Grey lines: $d^{-1.5}$ (solid line) uniform (dashed line) radial distributions used for SDO models. Black lines: $d^{-1.5}$ distribution (solid line) for SLOs, and intrinsic radial distribution (dashed line) of object with Sedna's $a$ and $e$. Vertical bars: Distance limits for characterized surveys (widths represent variation of limit with $i$). Vertical dotted line: Estimated distance limit for B04.}
\label{radial}
\end{centering}
\end{figure*}

%%% Figure 1 here

With the characterized outer limits of several literature pencil-beam surveys for TNOs, we can explore the upper limits these surveys put on distant populations. Few detections from these surveys are followed-up after enough time has elapsed for orbital properties other than heliocentric distance and inclination to be measured with any precision. With these short-arc measurements, it is difficult to identify the population any individual object belongs to unless it can be constrained by $d$ and $i$ alone. By determining the most distant detection in each survey under consideration and simulating a population that spends at least some fraction of time outside of this most distant detection, we can determine an upper limit for the size of this population based on the \textit{non}-detection of any sources at greater distances. Two such distant populations are the SDOs and SLOs.

We simulate each population by sampling objects from distributions of heliocentric distance $d$ and absolute magnitude $H$. For our model populations, we sample $H$ from a power-law absolute magnitude cumulative-distribution function (CDF) with the form 

\begin{equation}
N(\leq H) = N_{\leq H_{max}} 10^{\alpha(H-H_{max})},
\end{equation}

where $N_{\leq H_{max}}$ is scaled to produce the total number of objects in our sample, given $H_{max}$ is the maximum absolute magnitude of interest. 

Because the magnitude distributions of TNOs are thought to break to a lower slope at a characteristic magnitude and therefore depart from a true power-law, we will model our populations with magnitudes brighter than $H\sim10$ to avoid the putative break. Since we do not know the slope of the magnitude distributions \textit{a priori}, we will test our models over a plausible range of $\alpha$. 

For SDOs, heliocentric distance is sampled from two distributions. The first is a differential radial distribution with the form

\begin{equation}\label{dist}
\mbox{d} \! N \propto d^{-1.5} \, \mbox{d} \! d
\end{equation}

with the exponent of $-1.5$ as used in T00 when performing similar estimates of distant populations. In the T00 models, synthetic SDOs were generated over distances of $34-366$ AU. For our estimate of the upper limit of this population, we use the same distance range. The second distribution is uniform in heliocentric distance over the same distance range, in line with the uniform distribution measured by Kavelaars et al. (2008) and returned by the dynamical simulations of Volk \& Malhotra (2008). 

We also model the SLO population with two radial distributions. The first is the same radial distribution defined in Eqn. \ref{dist}, but over $70-1000$ AU. The second distribution is similar to that used by S09, which is the intrinsic radial distribution of an orbit with Sedna's $a$ and $e$. The radial distributions used by our simulations are illustrated in Figure \ref{radial}, with the outer distance limits of the surveys under consideration marked. 

We generate a large number of objects drawn simultaneously from each distribution and for each object determine its apparent magnitude $m \simeq H + 10\log(d)$. For each survey, we then determine the following:

\begin{enumerate}
\item Is the object within the heliocentric distance limit of the survey?
\item Given the object's apparent magnitude $m$, a uniform random probability $p$, and the survey's efficiency function $\eta(m)$ (clipped at 15\%), is the object detected ($p\leq\eta(m)$)?
\end{enumerate}

If an object satisfies both conditions, it is counted as ``detected'' by that survey. We also note whether or not the detection is outside the most distant $H\leq10$ detection made by the survey. We then scale the number of detections in each survey by the fractional area $f_{\Omega}$ of the entire sky the survey searched. Given this scaled detection count for each survey, we determine the mutual probability of non-detection of the synthetic population more distant than the most distant real detection for all surveys. We then determine the initial number of objects in the population required to return a mutual probability of non-detection $0.27\%$ ($3\sigma$ upper limit), given this population model. For every $\alpha$ we simulate, we take the corresponding derived upper limit on the population ($N(\leq H = 10$)) and convert it to $N(D \leq 100$ km) by assuming a 4\% albedo.

By making detection a function only of $H$ and $d$, we are effectively modeling a population that is 
isotropic on the sky, which implies an inclination distribution of $p(i) \propto \sin(i)$. 
 For a population with an unknown inclination distribution this is as reasonable an assumption as any, 
 and it significantly simplifies analysis. We note that T00 only considered inclinations as high as $35^\circ$ 
 for SDOs, and S09 modeled SLOs with an inclination probability distribution that was effectively 0 for $i \gtrsim 35^\circ$. 
 In order to compare to the earlier estimates on the upper limits of the SDO and SLO populations, we scale $f_{\Omega}$ 
 in our simulations by a factor of $1/ \sin(35^\circ)$, which has the effect of limiting the total area of the sky considered to that 
 which inclinations $ < 35^\circ$ can populate. 

Since the field locations of the pencil-beam surveys considered here all have ecliptic latitudes near $\sim0^\circ$, the upper limits we derive only truly reflect an extrapolation of the population density on the ecliptic. In order to determine the total population limit $N'$ implied by our surveys given an arbitrary latitude distribution, one may simply multiply our upper limit $N_{0}$ by the integral of the desired latitude distribution $P'(l)$ (where $P'(0^\circ)$ is normalized to 1) between $l$ of $0$ to $\pi/2$ radians, and then divide by the integral of our assumed distribution ($P_{0}(l) = \cos(l)$ between $0$ and $0.611$ radians, 0 elsewhere) over the same range: 
\begin{equation*}
N' = \frac{N_{0}}{0.57} \int\limits^{\pi/2}_{0} \! P'(l) dl . 
\end{equation*}
For example, assuming a latitude distribution for the SDOs similar to that determined by Brown (2001; approximated here as a gaussian with $\sigma=0.31$ radians) results in a correction of our upper limits by a factor of $\sim0.6-0.7$.

\section{Scattered Disk population limits}
%%% Figure 2 here

%%% Figure 2
\begin{figure*}
\begin{centering}
\includegraphics[width=10cm]{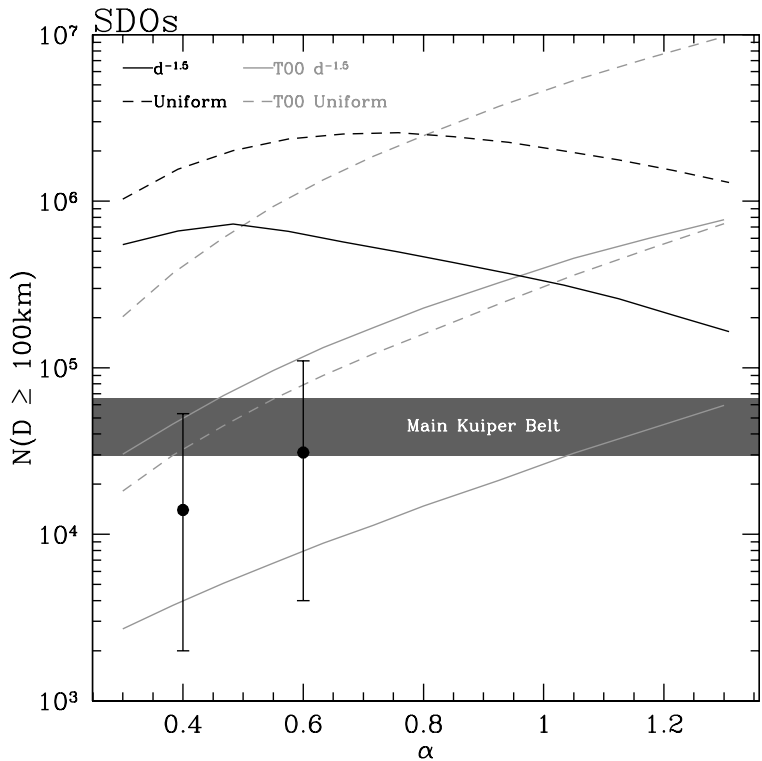}
\caption{Limits on total population of $i < 35^\circ$ Scattered-Disk Objects with diameters $\leq 100$ km vs. CDF slope $\alpha$. Solid lines: $3\sigma$ upper limit from pencil-beam surveys considered here (black) and extrapolated upper- and lower-limits from the detections in T00 (light gray) using the $d^{-1.5}$ radial distribution. 
The inflection point in the black line occurs at the value of $\alpha$ where the upper limit constraint on the population of SDOs transitions from being driven by detections at small distance ($\alpha>0.5$) to being driven by the lack of detections at large distance ($\alpha < 0.5$). 
Dashed lines: $3\sigma$ upper limit from pencil-beam surveys considered here (black) and extrapolated upper- and lower-limits from the detections in T00 (light gray) using a uniform radial distribution. Points: limits estimated by T00, with $3\sigma$ limits marked by error bars. Gray band: $3\sigma$ limits on Main Classical Kuiper Belt population by ``nominal'' model of K09, with $\alpha_{MKB} = 0.72$. }
\label{SDO}
\end{centering}
\end{figure*}

Since the perihelia of the SDO population extend well inside the outermost detections of the surveys considered here, the non-detection of distant sources is not necessarily a stronger constraint on the population than the total number of detections consistent with being SDOs in the survey. To account for this, we also count the number ($N_{sim}$) of $H\leq10$ simulated detections returned by each survey where the distance at detection of the simulated object was less than the distance of any real detection by that survey. We assume (as a limiting case) that \textit{all} $N_{real}$ detections with $H\leq10$ and $d > 34$ AU made by each survey were SDOs\footnote{Because these surveys do not provide accurate measurement of orbital parameters other than heliocentric distance and inclination, no more refined characterization of individual objects is possible.}. We then determine the probability of detecting $N_{real}$ given $N_{sim}$, and derive an upper limit when $N_{sim}$ is ruled out at the $3\sigma$ level. If this upper limit is stronger than the upper limit derived by the non-detection of distant objects for a given $\alpha$, we adopt it as our upper limit for that slope. 

In addition to modeling the pencil-beam surveys, we have derived separate limits on the SDO population by passing the detections of T00 through our simulation. T00 had several verified real detections, which allow us to derive both upper and lower limits for the population. Our upper limits in the $d^{-1.5}$ radial distribution case match those estimated by T00 well, while our lower limits are a factor of $\sim2$ higher than the values quoted in T00. This discrepancy is likely due to variations between the uniform inclination distribution used by T00 
and the uniform latitude distribution used here. 

Figure \ref{SDO} illustrates the results of our simulations for the $i < 35^\circ$ SDO population, the SDO limits extrapolated from our simulations of the detections in T00, and limits on the Main Kuiper Belt (MKB) population from the ``nominal'' model of Kavelaars et al. (2009; hereafter K09). For increasing $\alpha$, the pencil-beam survey upper limit on the population decreases, while the upper limit from T00 increases.  For both radial distribution models, the point where the two limits cross defines an absolute upper limit on the population for any CDF power-law slope $\alpha$ (given that radial distribution model). For the $d^{-1.5}$ radial distribution model, this crossover occurs at approximately $\alpha = 0.95$, with a maximum population size of $N(D>100$ km$) \leq 3.5\times10^5$ objects. For the uniform distribution model, this crossover occurs at the shallower slope of approximately $\alpha = 0.8$, with a maximum population size limited to $N(D>100$ km$) \leq 2.5\times10^6$ objects. These limits translate into $5-40$ times the upper limit population of the K09 ``nominal'' model of the MKB.

\section{Sedna-like population limits}

%%% Figure 3 here

%%% Figure 3
\begin{figure*}
\begin{centering}
\includegraphics[width=10cm]{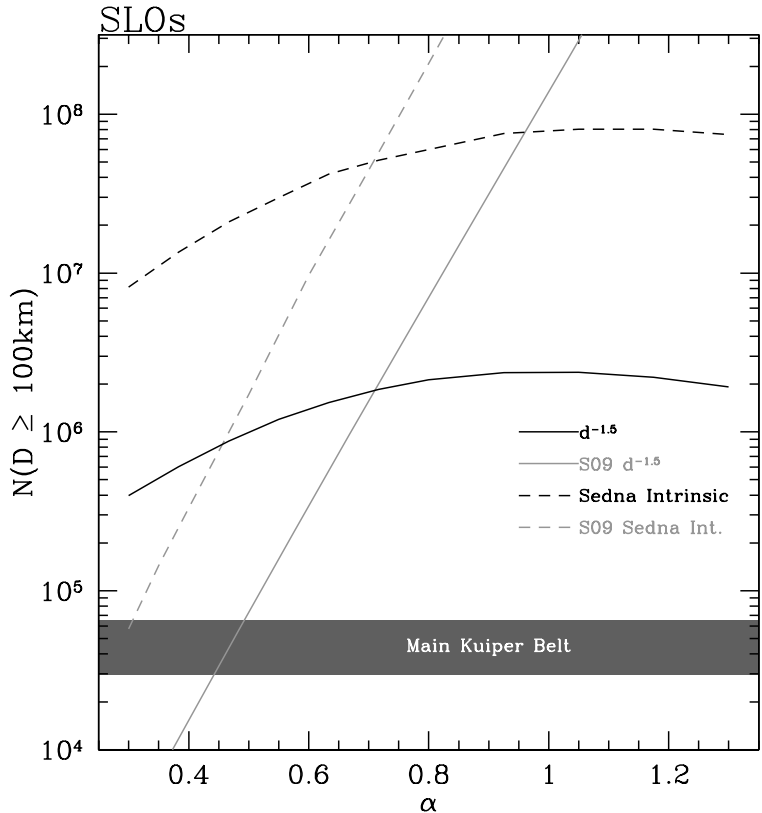}
\caption{Limits on total $i < 35^\circ$ Sedna-like population with diameters $\leq 100$ km vs. CDF slope $\alpha$. Solid lines: $3\sigma$ upper limit from pencil-beam surveys considered here (black) and extrapolated from S09 (light gray) using the $d^{-1.5}$ radial distribution. Dashed lines: $3\sigma$ upper limit from pencil-beam surveys considered here (black) and extrapolated from S09 (light gray) using Sedna's intrinsic radial distribution. Gray band: $3\sigma$ limits on Main Classical Kuiper Belt population by ``nominal'' model of K09, with $\alpha_{MKB} = 0.72$.}
\label{SEDNA}
\end{centering}
\end{figure*}

The most distant detections ($\sim43-65$ AU) in each of the pencil-surveys considered here are all interior to the modeled perihelia of the SLO populations ($70-76$ AU). The limiting magnitudes of these surveys allowed them to detect objects as faint as $H \simeq 10$ (B04 limit at 70 AU). While these surveys searched over 3,000 times less area than S09, their flux limits are up to nearly seven magnitudes fainter than the limit of S09 at perihelion, permitting a significant improvement on the limits for steep CDF slopes.

Comparing the SLO population estimates of S09 to the limits derived in this work is not straightforward. S09 was sensitive only to the largest objects in the population while the surveys analyzed in this work were sensitive to much smaller objects. We have modeled the detection of Sedna by S09 with the same populations used to simulate the pencil-beam surveys (with both radial distribution models) and extrapolated the $3\sigma$ upper-limits the S09 survey put on the bright population to diameters as small as 100 km. 

Figure \ref{SEDNA} illustrates the results of our simulations for the SLO population (given both radial distribution models), the SLO limits we derive from the detection of Sedna by S09 extrapolated to the $D\sim100$ km regime, and limits on the MKB population from the ``nominal'' model of K09. The large difference (factor of $\sim50$) between the two limits illustrates the sensitivity to assumed radial distribution models. 

The crossover where the upper limit from pencil-beam surveys becomes a stronger constraint on the SLO population is approximately $\alpha \simeq 0.7$ for the $d^{-1.5}$ radial distribution and $\alpha \simeq 0.63$ for the Sedna-like radial distribution. At the slope where the limit derived here is stronger than the extrapolated S09 result for each radial distribution model, our upper limit on the SLO population is $\sim10$ times larger than the K09 upper limit on the MKB for the $d^{-1.5}$ distribution, and the largest the SLO population can be for any CDF slope with this radial distribution is roughly 40 times the MKB upper limit. 
Given the Sedna-like radial distributions, the SLO population can be up to $\sim$1300 times more populous than 
the upper limit on the MKB (with the peak occurring at $\alpha \sim 1$) before the lack of detections in current 
pencil-beam surveys would be incongruous with expectations.

\subsection{Requirements for Future SLO Surveys}

%%% Figure 4
\begin{figure}
\begin{centering}
\includegraphics[width=8cm]{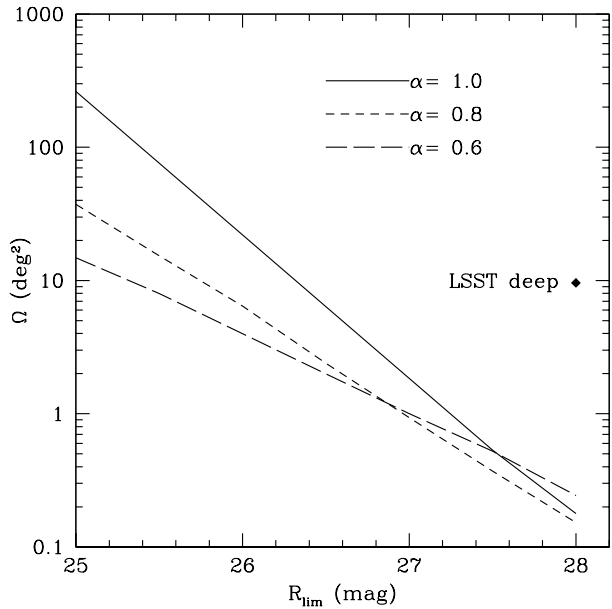}
\caption{Required survey area $\Omega$ (square degrees) vs. survey $R$ limiting magnitude in order to obtain stronger 3$\sigma$ upper limits on Sedna population than derived here, for magnitude-distribution power-law slopes $\alpha =$ 0.6, 0.8, and 1.0. Diamond: expected depth and area of a single LSST ``deep drilling'' field.}
\label{omega}
\end{centering}
\end{figure}

If a survey is undertaken to detect the SLO population, or at least drive down the upper limit on the population, how deep and over what area must the survey search? Given the uncertainty in the properties of the population, we have estimated the area $\Omega$ a survey must search, given a limiting magnitude $R_{\mbox{lim}}$ (and $\alpha =$ 0.6, 0.8, and 1.0), in order to obtain stronger $3\sigma$ upper limits on the SLO population than those derived here. Figure \ref{omega} illustrates the results for survey limiting magnitudes of $R_{\mbox{lim}} = 25-28$. The model survey is assumed to be sensitive to motions out to 1000 AU. A single Large Synoptic Survey Telescope (LSST) ``deep drilling'' field (Chesley et al. 2009) will probe to nearly the same depth as Bernstein et al. (2004) but over 500 times the area. If the methods used to search this component of the LSST data for TNOs is sensitized to distant objects, the chances for finally obtaining a statistically useful sample of the Sedna-like population are very good.

\section{Summary}

We have used the re-characterization by Parker \& Kavelaars (2010) of the orbital sensitivity of existing pencil-beam surveys in literature to derive new upper limits on distant populations of minor bodies in the Solar System. P10 considered three surveys (Fraser et al. 2008; Fraser \& Kavelaars 2009; and Fuentes et al. 2009) and found that for each survey the outer limit was significantly more distant than claimed in the original publication. Using these newly-derived outer limits, we determined the constraints these surveys (plus Bernstein et al. 2004) put on the Scattered Disk and Sedna-like populations. 

The combination of survey types allows us to limit the maximum possible size of the population for any CDF slope. 
Our limits constrain the Scattered Disk to be less than $5-40$ times more populous than the Main Kuiper Belt, 
depending on the radial distribution of material. For most plausible CDF slopes, SDOs are better constrained by 
relatively shallow, wide area surveys like those presented in Trujillo et al. (2000). For both a $d^{-1.5}$ and 
a uniform radial distribution model, we estimate the maximum population size at any CDF slope to less than 
$N(D>100$ km$) \leq 3.5 - 25\times10^5$, respectively.  

A strong sensitivity to the assumed radial distribution is demonstrated by the Sedna-like population, as up to 
$1300$ times the $D<100$ km Main Kuiper Belt population could reside on Sedna-like orbits (as assumed by the limits 
presented in S09) and still remain undetected in existing surveys, whereas if the material follows a $d^{-1.5}$ 
radial distribution this upper limit is reduced to 40 times the current population of the Main Kuiper Belt. 

At present the upper-limits on the population size determined by lack of detection in existing pencil-beam surveys do not appear 
to conflict with the population estimates based on wide-area, shallow surveys.  
This implies that the lack of distant detections in current-generation surveys are not indicative of any incongruences 
in survey sensitivity calibrations or other issues.

\section{Acknowledgements}

Alex H. Parker is supported by the NSF-GRFP award DGE-0836694. We thank our anonymous reviewer for their constructive comments which helped us clarify the description of our survey simulation method.

\end{document}